\documentclass[preprints,review,accept,moreauthors,pdflatex]{mdpi}

\firstpage{1} 
\pubvolume{xx}
\issuenum{1}
\articlenumber{5}
\pubyear{2019}
\copyrightyear{2019}
\history{Received: date; Accepted: date; Published: date}

\continuouspages{yes}
\usepackage{graphicx}	
\usepackage{amsmath}	
\usepackage{amssymb}	

\pdfoutput=1

\Title{The self-control of Cosmic Rays}

\newcommand{\cF}{{\cal F}} 
\newcommand{\be}{\begin{equation}} 
\newcommand{\ee}{\end{equation}} 
\Author{Pasquale Blasi $^{1,2}$\orcidA{0000-0003-2480-599X}}

\AuthorNames{Pasquale Blasi}

\address{%
$^1$Gran Sasso Science Institute, Via M. Iacobucci, 2, 67100 L'Aquila, Italy; pasquale.blasi@gssi.it\\
$^2$INFN/Laboratori Nazionali del Gran Sasso, Via G. Acitelli 22, Assergi (AQ), Italy}

\abstract{Several independent pieces of information have recently hinted at a prominent role of cosmic rays in controlling their own transport, within and around the sources as well as throughout their propagation on Galactic scales and even possibly during their escape from the Galaxy. I will discuss this topic with special attention to the theoretical implications and possible additional observational evidence that we may seek with upcoming experiments.}


\begin{document}


\section{Introduction}

The bulk of cosmic rays (CRs) is made of fully ionized light nuclei, moving at speeds close to the speed of light. In all cases of astrophysical interest, these particles propagate diffusively through magnetized media, with the possible exception of CRs at the highest energies, which might propagate quasi-ballistically. The diffusive ansatz seems appropriate to both acceleration regions and transport through the Galaxy. Yet, many aspects of this ansatz are somewhat less clear than one would think. Here I will list just some of the issues that arise when one attempts to write down a theory of CR diffusion: 1) the well established theory of resonant scattering off Alfv\'en waves (see \cite{BlasiRev2013} for a recent review, and references therein) describes well the diffusive transport parallel to the ordered magnetic field, but it is all but clear that the escape of CRs from the Galaxy is actually related to such parallel diffusion, as discussed from the very beginning \cite{1969ApJ...155..777J,1969ApJ...155..799J}; 2) for quite some time, the nature of the Alfv\'en waves responsible for CR scattering has been associated with some kind of turbulent cascading from large scales, but it has been recognized that such cascade proceeds anisotropically, favoring the perpendicular direction \cite{GS1,GS2}, so as to make effective resonant scattering hard to achieve; 3) diffusive motion close to shock fronts, most notably associated to supernova explosions, is usually considered as a crucial ingredient of diffusive shock acceleration, but as first recognized by \cite{Lagage1,Lagage2}, the interstellar turbulence can only account for acceleration up to a few GeV, orders of magnitude short of the observed CR energies. Hence CRs have long been suspected of generating waves upstream of shocks so as to shorten the acceleration time and increase the maximum achievable energy. Progress keeps happening in this area (see for instance \cite{Bell2004}); 4) The best fit to the low energy part of the Boron/Carbon ratio seems to require either effective CR reacceleration on plasma waves or steep energy dependence of the diffusion coefficient. It has been argued that the reacceleration scenario might run into energetic problems \cite{DruryReac} and the steep energy dependence of the diffusion coefficient should result in excessive large scale anisotropy \cite{PtuskinAni,AmatoBlasiAni}. 

To make the situation even more interesting, recent data collected by PAMELA \cite{PAMELA-P-He} and AMS-02 \cite{AMS-P,AMS-He}, show that the spectra of CR protons and helium nuclei are characterized by a break at rigidity of $\sim 200-300$ GV, which most likely reflects a change in the transport properties of CRs \cite{Genolini}. AMS-02 has recently showed that a similar break exists in the spectra of all primary nuclei \cite{AMS-nuclei}. Moreover, the recent measurements carried out by AMS-02 of the spectra of secondary elements such as boron, lithium and beryllium \cite{AMS-Sec} shows that their hardening is stronger than for primaries, thereby supporting the hypothesis that the phenomenon is associated with CR transport rather than with the accidental proximity of a source to the solar system or to phenomena related to acceleration in the sources. 

All these results stimulated an enhanced interest in the investigation of the physics of CR transport in the Galaxy and have led to several interesting results put forward in the last few years. One of the most interesting topics in this context is the self-controlling action of CRs, namely their role in changing the environment in which they propagate, both inside the acceleration region and during Galactic transport, so that the diffusive properties are no longer independent of CRs themselves but rather determined by them. This interaction is mediated by the excitation of plasma instabilities associated with the current of CRs that develops in any situation in which CRs are not homogeneously distributed in space (which reflects in local anisotropy). Depending on the conditions, different types of instabilities may develop and result in a diverse range of interactions between CRs and the environment in which they move. This phenomenon is generally referred to as non-linear transport and will be the main focus of this article. 

The paper is organized as follows:  In \S \ref{sec:GalTra} I will emphasize the effects of the excitation of streaming instability on Galactic scales, as due to the global gradient of CRs escaping the Galaxy. Such gradients may be much larger, though time dependent, close to sources, while accelerated particles are turning into CRs. The enhanced self-confinement close to sources, such as supernova remnants (SNRs), will be described in \S \ref{sec:sources}. In \S \ref{sec:escape} I will describe the implications of self-excitation of non resonant streaming instability on the escape of CRs from the Galaxy and the challenges that these findings pose for the interpretation of CR escape into the intergalactic medium. I will summarize these topics and highlight future developments in \S \ref{sec:conclude}.

\section{Non-linear effects in CR Galactic transport}\label{sec:GalTra}

The role of CRs in creating their own scattering centers was first highlighted immediately after the theory of CR transport in hydromagnetic waves was developed \cite{1969ApJ...156..303W,1969ApJ...156..445K,Skilling75,Holmes}. The gradient of CRs inside the Galaxy can be easily calculated in a simple model of diffusive transport: if particles escape freely at the edge of the magnetized halo of size $H$, and $f_{0}(p)$ is the phase space particle distribution function at the disc of the Galaxy (assumed to be infinitely thin), the gradient is of the order of $f_{0}(p)/H$. This induces a diffusive current of CRs directed outward:
\be
J_{CR}(p)=4\pi p^{3} e f_{0}(p) v_{D} = e D(p) 4\pi p^{3} \frac{f_{0}}{H},
\ee
where $v_{D} = D/H$ is the drift velocity. When CRs drift at speed larger than the Alfv\'en speed $v_{A}$, a streaming instability is excited on scales $k^{-1}$ that are resonant with the Larmor gyration of the particles making the current. The instability of modes with wavenumber $k$ grow at a rate that is proportional to the CR gradient calculated at the resonant momentum $p=q B_{0}/kc$:
\be
\Gamma_{\rm CR}(k) = 
\frac{16 \pi^{2}}{3} \frac{v_{\rm A}}{\cF B_{0}^{2}} \left[ p^{4} v(p) \frac{\partial f}{\partial z}\right]_{p=q B_{0}/kc},
\label{eq:gammacr}
\ee 
where $\cF (k)$ is the dimensionless power in modes of wavenumber $k$ per unit logarithmic scale. 

There are two algebraic handlings of Eq. \ref{eq:gammacr} that help clarifying its physical relevance. We first notice that the quasi-linear expression for the diffusion coefficient (parallel to magnetic field lines) can be written as:
\be
D(p) \approx \frac{1}{3}r_{L}(p) v(p) \frac{1}{\cF (k_{res})},
\ee
where $k_{res}=1/r_{L}(p)$. After a few steps, one can rewrite Eq. \ref{eq:gammacr} in the form:
\be
\Gamma_{\rm CR}(k)  \approx \frac{v_{D}}{v_{A}} \Omega_{c} \frac{n_{0}(>p)}{n_{i}}~~~~~~~v_{D}>v_{A},
\label{eq:simple1}
\ee
where $n_{0}(>p)=4\pi p^{3} f_{0} (p)$ is the number density of CRs with momentum $>p$, $n_{i}$ is the gas density and we introduced the cyclotron frequency $\Omega_{c}=e B_{0}/m_{p} c$. A few considerations are in order: the best fit to the transport of CRs in the Galaxy made with very diverse approaches return the same basic conclusion, namely that assuming a diffusion coefficient in the form $D(R)=D_{0}(R/3 GV)^{\delta}\rm cm^{2}/s$, data are best described by adopting $D_{0}/H\sim 1$ if $D_{0}$ is in units of $10^{28}$ and $H$ is in units of $kpc$. Keeping in mind the discussion above, this translates to a minimum drift velocity $v_{D}=D/H\sim 3\times 10^{6} (R/3 GV)^{\delta}\rm cm/s$. Adopting a gas density in the halo (where CRs spend most of the time) of $n_{i}\approx 10^{-2}\rm cm^{-3}$ and a magnetic field $B_{0}\sim 3\mu G$, the Alfv\'en speed is $v_{A}=6\times 10^{6}\rm cm/s$, which implies that $v_{D}>v_{A}$ for rigidities $R>10$GV for typical values $\delta\sim 1/3-1/2$. Using the observed density of CRs at 10 GV as measured by AMS-02 \cite{AMS-P} in Eq. \ref{eq:simple1} one obtains a typical growth time for CRs of rigidity 10 GV of about one thousand years, very short compared with the typical escape time from the Galaxy, as inferred from the measurements of secondary/primary ratios. It follows that the phenomenon is potentially very important for high energy CRs.

Let us look at Eq. \ref{eq:gammacr} from another point of view: if we introduce the magnetic energy density $U_{B}=B_{0}^{2}/4\pi$ and the CR pressure $P_{cr}(>p) = \frac{4\pi}{3}p^{4} v f_{0}(p)$, one can rewrite Eq. \ref{eq:gammacr} as follows:
\be
\Gamma_{\rm CR}(k) = \frac{P_{CR}(>p)}{U_{B}}\frac{v_{A}}{H \cF (k)}.
\label{eq:simple2}
\ee
The instability is usually assumed to be quenched by the onset of some type of damping mechanism. If the medium where CRs propagate were filled with neutral hydrogen, Alfv\'en waves would be mainly damped due to ion-neutral damping \cite{1969ApJ...156..445K} and the damping rate is so high that diffusive transport would be hard to justify. On the other hand, the standard picture of CR Galactic transport is based on the existence of a large, under-dense, ionized gas halo, with size of several kpc, where CRs spend most time before escaping the Galaxy. In such halo the main damping process is expected to be non-linear Landau damping \cite{LeeVolk} (NLLD), which proceeds at a rate:
\be
\Gamma_{NLL}(k) \approx k v_{s} \cF(k),
\label{eq:NLL}
\ee
where $v_{s}\simeq 3\times 10^{6}\rm cm/s$ is the sound speed. The saturation of the streaming instability occurs when the rate of growth equals the rate of damping, a condition that allows to estimate the power spectrum $\cF (k)$. Imposing equality between Eq. \ref{eq:simple2} and Eq. \ref{eq:NLL} one gets:
\be
\cF (k) = \left[ \frac{P_{CR}(>p)}{U_{B}} \frac{v_{A}}{v_{s}} \frac{r_{L}(p)}{H}\right]^{1/2},
\label{eq:Fk}
\ee
where we imposed the resonance condition $k=1/r_{L}(p)$. For the sake of simplicity let us adopt the expression $P_{CR}(>p)\approx 8.3\times 10^{-2}(E/10 GeV)^{-0.7}~\rm eV cm^{-3}$ for the CR pressure normalized at 10 GeV and $U_{B}=0.44~\rm eV cm^{-3}$ for the magnetic energy density corresponding to $B_{0}=3\mu G$. Since we will be dealing with relativistic CR particles, in the following we will use energy $E$, momentum $p$ and wavenumber $k=1/r_{L}(p)$ as referring to the same physical quantity. Hence, from Eq. \ref{eq:Fk}:
\be
\cF (k) \approx 1.8\times 10^{-5} \left( \frac{E}{10 GeV}\right)^{0.15} \left( \frac{H}{4 kpc}\right)^{-1/2} ,
\ee
and using the quasi-linear expression for the parallel diffusion coefficient (justified by the fact that $\cF\ll 1$), one gets:
\be
D(E) \approx \frac{1}{3} r_{L} v \frac{1}{\cF (E)} = 6\times 10^{27} \left( \frac{E}{10 GeV}\right)^{0.85} \left( \frac{H}{4 kpc}\right)^{1/2} ~ \rm cm^{2} s^{-1}.
\ee
Interestingly, this estimate of the self-generated diffusion coefficient agrees, within a factor $\sim 2$ with the diffusion coefficient inferred from phenomenological fits to CR and gamma ray data, suggesting that streaming instability is likely crucial in determining the scattering properties of CRs in the Galaxy. On the other hand, the energy dependence derived from imposing the balance between wave growth and NLLD seems to be too steep, although in more realistic models this slope does not reflect directly onto observable quantities such as the secondary/primary ratios (see for instance Ref. \cite{AloisioBlasi}).

The above discussion highlights the crucial role of damping processes for the determination of the saturation of streaming instability. The NLLD rate given in Eq. \ref{eq:NLL} was calculated in Ref. \cite{LeeVolk} and other articles as well, but a critical assessment of the role of this mechanism, perhaps using PIC simulations, is still missing. In fact, a change in the functional form of this damping rate might lead to a substantially different normalization and energy dependence of the diffusion coefficient of CRs. Interestingly however, both the inclusion of NLLD and turbulent cascading seem to imply a very similar power spectrum of the excited waves, and hence similar diffusion coefficients \cite{BlasiAmatoSerpico}, close to the one derived from phenomenological fits to observations. 

Magnetic fluctuations in the Galaxy are also produced by other phenomena, not directly related to CR streaming: for instance supernova explosions and stellar winds also stir the interstellar plasma on large scales. Such perturbations cascade to smaller spatial scales as it is usually the case for turbulence. The competition between perturbations that are self-generated through CR streaming and cascading turbulence has been subject of dedicated investigation \cite{BlasiAmatoSerpico,AloisioBlasi,AloisioBlasiSerpico}. However, magnetic perturbations cascade toward smaller scales while becoming anisotropic \cite{GS1,GS2}, which makes the problem of particle scattering rather tricky: the power in the modes parallel to the local magnetic field appears to be too small to justify the resonant scattering at the level needed to interpret CR observations of stable secondary nuclei. 

Aside from these unclear aspects, it is interesting that the shape of the power spectrum of perturbations generated through cascading and through CR streaming instability have different $k$-dependences, hence a transition between the two regimes should be expected, as also discussed in Ref. \cite{Farmer}. In \cite{BlasiAmatoSerpico} it was argued that the transition should appear in the form of a change in the energy dependence of the diffusion coefficient for CR energies around $\sim TeV$, which in turn reflects into a CR spectral break. Such breaks have been measured by the PAMELA \cite{PAMELA-P-He} and AMS-02 \cite{AMS-P} experiments and are commonly attributed to new phenomena associated with CR transport \cite{Genolini}. The recent measurements of the spectra of stable secondary CR nuclei, such as boron and lithium, by the AMS-02 experiment have confirmed that these features are most likely connected with a change of regime of CR transport at rigidity $\sim 300$ GV \cite{AMS-Sec}. On the other hand, whether this change is to be attributed to the onset of non-linear CR transport or more mondane explanations remains to be seen. For instance, it has been proposed that a diffusion coefficient with energy dependence that is a function of the height above the Galactic disc also leads to a CR spectrum at the Earth with a break in the TV rigidity range \cite{Tomassetti}, provided that the spatial and energy dependences are suitably chosen. In this sense, the attractiveness of the model based on non-linear effects in CR transport is based on the fact that the spectral break is expected in the right energy region without much freedom of choice of the parameters. It should be emphasized that the effect of waves excited in the Galactic disc and advected away from the disc and into the halo naturally leads to a spatial dependence of the diffusion coefficient \cite{Carmelo} and these models also provide a good description of the data, combining non-linear effects and the consequent spatial dependence of the diffusion coefficient. 

The production of waves able to scatter CRs resonantly have also been invoked in models of CR driven winds, where the waves provide the coupling between CRs and the interstellar gas. The launching of an outflow (in special conditions even a wind) depends on the local balance between the pressure exerted by CRs and the gravitational pull dominated by the action of dark matter. A magnetohydrodynamic semi-analytic approach to these CR driven winds was first developed in Ref. \cite{PtuskinWind1}, where CRs were treated as a fluid with given diffusivity and the geometry of the outflow was fixed based on a physical picture of a 1D outflow. By construction these approaches lack information about the spectrum of CRs which is affected by the presence of the wind both because of the different coupling with waves (diffusion) and because of the advection with the wind (and the waves), unavoidable in CR driven winds. A recipe to infer the asymptotic (low energy and high energy) shape of the CR spectrum in the Galactic disc was put forward in \cite{PtuskinWind2}. More recently, MHD simulations of the development of CR driven winds have been developed \cite{Pfrommer1,Pfrommer2,Pfrommer3}, in which a realistic geometry of the outflow is derived. However, such approaches either treat CRs as a fluid, thereby lacking information about their spectrum, or assume a given diffusion coefficient, thereby lacking the treatment of particle transport in self-generated waves. 

The equations of conservation of mass, momentum and energy including CRs and waves generated through streaming instability have recently been solved together with the CR transport equation \cite{Recchia1,Recchia2}. This approach showed how most of the wind solutions that may be found in fact lead to CR spectra in the Galactic disc that are quite unlike the ones observed at the Earth. It is likely that in Nature winds are launched preferentially in some regions of the Galaxy rather than from the whole disc. This would however lead to the conclusion that the CR spectrum may be quite different in different parts of the Galaxy. It remains to be seen whether such scenarios are compatible with the remarkable homogeneity of the CR distribution that gamma ray observations show. 

\section{From accelerated particles to CRs: non-linear effects around sources}\label{sec:sources}

Non-linear effects in CR transport are more prominent in the presence of large CR gradients, as can be inferred from Eq. \ref{eq:gammacr}. This condition is most easily verified in the proximity of Galactic CR sources, such as supernova remnants and clusters of young stars. The CR acceleration history of a SNR typically lasts for $\lesssim 10^{5}$ years, when the shock produced by the explosion enters the radiative phase.  In the case of stellar clusters, the CR production may be considered quasi-stationary if it is associated with collective acceleration in stellar winds and the lifetime of the massive stars in the clusters is of order $\sim 10^{7}$ years. 

The cloud of accelerated particles moving away from the source is mainly made of low energy particles, namely particles with gyroradius  much smaller than the coherence scale of the Galactic magnetic field in which the source is located, which is typically assumed to be $\sim L_{c}\sim 10-100$ pc. In fact the local field might retain its approximate direction on scales larger than $L_{c}$ if $\delta B/B_{0}\ll 1$. In this situation, diffusion occurs mainly in the direction of the ordered local field, and can be approximated as one dimensional. At distances larger than $L_{c}$, diffusion becomes three dimensional and eventually the cloud of particles originated at the source merges into the diffuse CR background. It is clear however that the CR density around the source may stay much larger than that of the background (diffuse Galactic CRs) for approximately a diffusion time. 

Assuming that the diffusion coefficient outside the source is of the same order as the one derived from secondary/primary ratios (for instance B/C), one concludes that the cloud of CRs stays close to the source (within a distance $L_{c}$) for a time:
\be
\tau \approx \frac{L_{c}^{2}}{D_{gal}(E)} \approx 2000 L_{c,10}^{2} E_{GeV}^{-\delta} ~ \rm years ,
\label{eq:conf}
\ee
where $L_{c,10}=L_{c}/10 pc$. The assumption of diffusive transport is clearly applicable only for energies such that $\tau\gg L_{c}/c$, namely $E\ll 200$ TeV $L_{c,10}^{3}$, where we used $\delta=1/3$ for the numerical estimate.  

The gradient established by CRs diffusing away from the source is large enough to induce streaming instability in the region around the source, as discussed in Refs. \cite{Plesser,Malkov,Dangelo,Nava,NavaRecchia}. The level of self-generated turbulence of these particles depends rather critically on the amount of neutral hydrogen present in the region around the source, because of the strong ion-neutral damping that is induced by such gas. If the assumption is made that the gas is fully ionized in a region of $50-100$ pc around the source, which typically corresponds to very low densities ($n\sim 10^{-2}-10^{-3}\rm cm^{-3}$), the effect of self-generated waves can be very important in lowering the diffusion coefficient and increasing the residence time in the same region to several hundred thousand years instead of the short time suggested by Eq. \ref{eq:conf}. However the grammage accumulated in the near source region is negligible, unless a dense, concentrated cloud is present in a small portion of the volume around the source. The effects of self-generation and damping on the grammage accumulated by CRs in the near source region have been discussed by \cite{Dangelo,Nava,NavaRecchia}.

The CR self-confinement has also implications on the intensity and morphology of the diffuse gamma ray emission from the Galactic disc. As discussed by \cite{DangeloGamma}, along some lines of sight the diffuse emission due to the overlap of numerous near source regions where diffusion is suppressed by the same particles may be comparable with the expected diffuse emission as derived in the assumption that CRs diffuse on Galactic scales. As mentioned above, the importance of these effects depends rather dramatically upon the presence of neutral hydrogen and/or the occasional presence of a dense localized cloud of gas in the near source region. The latter case may not be unusual since massive stars that type II SN explosions originate from are often located close to dense molecular clouds. 

One of the most important potential implications of the non-linear effects in CR diffusion in the near source regions is the modification that they imply on the global picture of CR diffusion in the Galaxy, especially in the aftermath of the recent AMS-02 observations of antiprotons \cite{AMS-antip} and positrons \cite{AMS-pos}. 

It was noticed \cite{Lipari} that 1) the spectrum of observed protons, positrons and antiprotons have very similar shapes, contrary to what is expected based on the standard picture of CR transport in the Galaxy. Moreover, 2) the ratio of fluxes of positrons and antiprotons is about the same as the ratio of their production cross sections. These two empirical facts were then interpreted as hints that both positrons and antiprotons could simply result from CR inelastic interactions in the interstellar medium, with no need for additional sources of positrons, such as pulsar wind nebulae (PWNe) \cite{BlasiPWN,PWNAmato} and old SNRs \cite{BlasiSNR}. These alternative scenarios (see \cite{alternative} for a review of conventional and alternative models of CR transport) represent radical changes to the pillars of the paradigm for the origin of Galactic CRs, as it is clear from the fact that the standard indicators of CR transport, such as the B/C ratio, has a behaviour that is qualitatively consistent with what expected based on the standard picture \cite{AMS-BC}, while positrons and antiprotons do not. A possible implementation of an alternative view of CR transport is the so-called {\it nested leaky box model} (NLB) \cite{Cowsik}, based on the idea that CRs may accumulate an energy dependent grammage around sources rather than on Galactic scales. The energy dependence is such that CRs with energy $> 100~\rm GeV$ escape the near-source regions in too short a time and end up accumulating grammage while propagating in the Galaxy, perhaps in an energy independent manner, as speculated in Ref. \cite{Cowsik}. 

These considerations are of crucial importance to understand why the $B/C$ ratio drops with energy as expected, while antiprotons and positrons behave differently. Boron nuclei from spallation reactions have the same energy per nucleon as the parent $C$ or $O$ nuclei, hence for energy per nucleon below $\sim 100$ GeV/n, one should expect the ratio to reflect the near source grammage, which in the NLB model is postulated to drop with energy. For a positron or an antiproton, the situation is different, in that they are typically produced in inelastic collisions of protons with energy $\sim 20$ times larger. A positron at $\gtrsim 10$ GeV is produced by a proton with energy $\gtrsim 200$ GeV, for which the grammage is dominated by the energy independent Galactic grammage, hence the spectrum of such secondary positrons (or antiproton) would be expected to be the same as that of protons. It must be clear that this class of alternative models is not problem free: they require that the role of energy losses is marginal for electrons and positrons with energy $\lesssim 1$ TeV, otherwise the positron spectrum is modified with respect to that of protons. Moreover the spectrum of protons that we observe at the Earth is required to be the same as that at the source, since the Galactic grammage is assumed to be energy independent. Finally electrons and protons in the sources, whatever they might be, are required to be different, so as to explain why the observed electron spectrum is different from that of protons. 

Despite these difficulties, these alternative models are attracting some attention. The analogy between the cocoons of the NLB model and the near source regions due to the non linear CR transport is tantalizing, hence some efforts are being devoted to understanding whether CRs may accumulate some level of grammage in such regions. In addition it is worth recalling that some recent observations of diffuse gamma ray emission around SNRs \cite{Fermi-SNR} and around PWNe \cite{2017Sci...358..911A} have provided clear evidence for a diffusion coefficient in such regions that is 10-100 times smaller than the one inferred from the measurements of the B/C ratio. In the case of PWNe the non linear effects induced by electron-positron pairs escaping the pulsar environment seem to be insufficient to explain the observations \cite{2018PhRvD..98f3017E}.

\section{Escape from the Galaxy}
\label{sec:escape}

The escape of CRs from our Galaxy (or from any galaxy for that matter) is as unclear as the escape from sources such as SNRs in our Galaxy. The spectrum of CRs at the Earth is typically calculated by solving the transport equation (including advection, diffusion and energy losses) with a free escape boundary condition at the edge of the halo. Such edge is inserted in the problem by hand and only sparse attempts to provide a physical explanation of it have been made through the years. 

In a basic approach to this problem, in which we only account for CR protons and we consider diffusion as the dominant process of CR transport, the current of CRs crossing the free escape boundary is given by $D\nabla n_{CR}(z=H)$ and this quantity equals the flux of particles injected at the disk through SN explosions. This current is only weakly model dependent and can be considered as a very reliable piece of information. In a one dimensional model with diffusion only, the escape flux can be easily calculated by integrating the transport equation, and reads:
\be
\phi_{CR}(E) = - D(E) \frac{\partial n_{gal}}{\partial z} = D\frac{n_{gal}}{H}= \frac{L_{CR}}{2\pi R_{d}^{2}\Lambda}E^{-2}\ ,
\label{eq:phi}
\ee
where we have assumed that CRs are injected in the Galactic disc of radius $R_{d}$, with a luminosity $L_{CR}$ and a spectrum $\propto E^{-2}$ extending between $E_{min}$ and $E_{max}$, and $\Lambda=\ln(E_{max}/E_{min})$. Eq.~\ref{eq:phi} clearly shows that, as expected, the spectrum of escaping CRs is the same as the injected spectrum. 

The condition of free escape mimics the ballistic motion of CRs outside the halo, which implies that the number density drops dramatically after escape, $n_{CR,ext}(E)=3\phi_{CR}/c$, where the ballistic motion is assumed to occur at the speed of light. Yet the flux of particles diffusing out toward the free escape boundary is exactly the same as the flux of particles moving ballistically away from the halo: the conserved current carried by CRs with energy $>E$, is then given by $J_{CR}=e E\phi_{CR}(E)$. The question arises of whether this current is able to perturb the circumgalactic medium and thereby affect the ballistic motion of CRs. This issue was addressed in detail in a recent work \cite{BlasiEscape}.

The magnetic field of the Galaxy is expected to drop down exponentially on spatial scales of a few kpc away from the Galactic disk. A qualitative expectation is that this drop would reflect in a corresponding increase of the diffusion coefficient. Such a trend implies that at some point the motion of the particles can be well approximated as ballistic: the location where this happens is, in good approximation, to be identified with the free escape boundary. The magnetic field eventually reaches the value corresponding to the intergalactic magnetic field $B_{igm}$, for which very weak upper limits exist \cite{Burles}. A lower limit, of order $\sim 10^{-17}$ G, has been recently inferred based on the non detection of low energy gamma rays from distant sources \cite{LowB}. The magnetic field in the circumgalactic medium, $B_{0}$, is expected to be somewhat higher than the intergalactic medium field because of the effect of gravitational collapse that gave rise to our Galaxy. 

As discussed in \cite{Bell2004} a non-resonant instability is induced by the current of CRs escaping the Galaxy, provided the following condition is fulfilled:
\be
\frac{E^{2}\phi_{CR}}{c} > \frac{B_{0}^{2}}{4\pi}.
\label{eq:nr}
\ee
The instability is excited on scales that are initially much smaller than the Larmor radius of the particles dominating the current, namely at wavenumber:
\be
k_{max} = \frac{4\pi}{c B_{0}}J_{CR}=\frac{4\pi}{cB_0^2}\frac{E^2\phi_{CR}}{r_L(E)}\ ,
\ee
and with a growth rate $\gamma_{max}=k_{max}v_{A}$, where $v_{A}=B_{0}/\sqrt{4\pi \rho}$ is the Alfv\'en speed in the unperturbed field and the density $\rho$ is written as $\delta_{G} \Omega_{b}\rho_{cr}$, where $\Omega_{b}$ is the baryon fraction in the universe and $\rho_{cr}$ is its critical density. The parameter $\delta_{G}\gtrsim1$ accounts for the overdensity of baryons around the Galaxy. 

The condition for the excitation of the non-resonant instability, Eq.~\ref{eq:nr}, translates into a condition on the background magnetic field:
\be
B_{0}\leq B_{sat}\approx 2.2 \times 10^{-8}  L_{41}^{1/2} R_{10}^{-1}\ {\rm G},
\label{eq:Bsat}
\ee
where $L_{41}$ is the CR luminosity of the Galaxy in units of $10^{41}$ erg s$^{-1}$ and $R_{10}$ is the radius of the galactic disk in units of 10 kpc.

When the instability is excited, its growth proceeds at a rate
\be
\gamma_{max}=k_{max}v_A\approx 2\ {\rm yr}^{-1}\ \delta_G^{-1/2} E_{\rm GeV}^{-1} L_{41} R_{10}^{-2}\ .
\ee
For values of $\delta_G$ appropriate to our Galaxy, the growth is extremely fast and the field rapidly grows. Finite temperature effects in the background gas might change this simple result, but, as discussed in Ref. \cite{BlasiEscape}, the main conclusions should not change for typical values of the parameters of the problem. 

Since $k_{max}^{-1}$ is initially much smaller than the Larmor radius of the particles dominating the current, the current is not much affected by the growth of the field. The Lorentz force $\sim J_{CR}\delta B/c$ displaces the background plasma by an amount $\Delta r \sim \delta B J_{CR}/c \rho \gamma_{max}^{2}$. The instability eventually saturates when the scale $\Delta r$ becomes of the same order of magnitude of the Larmor radius, which implies
\be
\delta B\approx B_{\rm sat}\approx\sqrt{\frac{2 L_{CR}}{c\ R_d^2\Lambda}}.
\label{eq:saturation}
\ee
It is worth stressing that this recipe for saturation might be complicated by numerous effects, such as thermal effects (see for instance \cite{zweibel10}). As discussed in Ref. \cite{BlasiEscape}, due to such effects, a dependence of the conclusions on the original field $B_{0}$ might appear, especially for very low values of $B_{0}$.

For a spectrum $N(E)\propto E^{-2}$, $\delta B$ is the same on all scales, so that the diffusion coefficient that results from the amplification processes is expected to be Bohm-like, $D(E)\propto E$. Moreover $\delta B$ is independent of the initial magnetic field strength and the density of background plasma. In \cite{BlasiEscape} the authors discuss how the small diffusion coefficient induced by this chain of phenomena leads to a CR density gradient that corresponds to a force active on the background plasma that is large enough to set it in motion with a velocity that is approximately the Alfv\'en speed in the amplified field $\tilde v_{A}=\delta B/(4\pi\rho)^{1/2}\sim 10-100$ km/s. The transport of CRs then becomeds advective and the CR density outside the halo can be written as:
\be
n_{CR}(E) = \frac{\phi_{CR}}{\tilde v_{A}}.
\ee 
The assumption of free escape of CRs from the Galaxy leads us to conclude that the strong instability induced by the CR current outside the halo brings the cloud of CRs to a stop and this sets the background plasma in motion with the local Alfv\'en speed in the amplified field. As a consequence, a large overdensity of order $\sim c/\tilde v_{A}\sim 10^{4}-10^{5}$ with respect to the case of free escape is established in a region of a few times $R_{d}$ around the Galaxy, where a magnetic field $\sim \delta B\sim 2\times 10^{-8}$ G is also created due to the action of CRs. The occasional interactions of CRs with the circumgalactic gas leads to production of gamma rays and neutrinos. The flux of neutrinos can be estimated as follows:
\be
F_{\nu}(E_{\nu}) E_{\nu}^{2} \approx  \frac{L_{CR}}{2 \pi R_d^2 \Lambda \tilde v_A}\frac{E_\nu^2}{E^2}
\frac{dE}{dE_{\nu}} \frac{\delta_{G} \Omega_{b} \rho_{cr}}{m_{p}} \frac{c \sigma_{pp} R_{d}}{2\pi}.
\ee

Calculating $\tilde v_A$ in the magnetic field $\delta B$ from Eq.~\ref{eq:saturation} we obtain
$$
F_{\nu}(E_{\nu}) E_{\nu}^{2} \approx \left(\frac{L_{CR}}{\Lambda}\right)^{\frac{1}{2}}\left(\frac{c \delta_{G}\Omega_{b}\rho_{cr}}{2\pi}\right)^{3/2}
\frac{\eta \sigma_{pp}}{m_p} = 
$$
\be
5\times 10^{-12} \delta_{G}^{3/2} \rm GeV cm^{-2} s^{-1} sr^{-1},
\ee
where we assumed that the neutrino energy is related to the energy of the parent proton by $E_{\nu}=\eta E$, with $\eta \sim 0.05$. For simplicity we neglected the weak energy dependence of the cross section for neutrino production, which is known to increase slowly with energy. The estimated flux of diffuse neutrinos depends on the overdensity $\delta_{G}$. If the overdensity of baryonic gas in the circumgalactic medium is of order $\sim 100$, then the expected neutrino flux is comparable with the one measured by IceCube \cite{Icecube1,Icecube2}. The same interactions also lead to the production of gamma rays through the decay of neutral pions. The comparison of such emission (including the effect of absorption due to pair production \cite{Ahlers}) with the Fermi diffuse gamma ray background \cite{Fermi-LAT} and with upper limits from other experiments at higher energies may lead to additional constraints on the required values of the overdensity $\delta_{G}$. It is worthwhile to mention that the virial radius of our Galaxy, which is of order $\sim 100$ kpc, is defined as the radius inside which the mean overdensity is 200. Hence a value of $\delta_{G}\sim 100-200$ appears to be quite well justified on scales of $\sim 10$ kpc. 

Models in which astrophysical neutrinos are assumed to be of Galactic origin have been previously put forward, with neutrinos produced in hadronic interactions in the halo \cite{Taylor}. Notice however that the CR transport in the halo is severely constrained by measurements of the secondary/primary ratios (such as B/C). It is not clear if the properties of CR transport invoked to explain neutrino fluxes in such models are compatible with the observed B/C ratio. This potential issue is absent in the scenario discussed above, in that CR transport in the region outside the halo does not affect the properties of CRs observed at the Earth.

It is worth keeping in mind that a physical process similar to that outlined above is expected to also take place around more luminous galaxies, potential sources of ultra high energy cosmic rays and leads to confinement of the lower energy part of such particles around the galaxies hosting the sources \cite{UHE}.

\section{Summary and Conclusions}
\label{sec:conclude}

CR diffusion is often described in phenomenological terms by introducing a diffusion coefficient (function of energy). Although useful, this approach misses the well known fact that CRs may generate their own scattering centers through the excitation of streaming instability. This instability, depending on the conditions, may proceed in a resonant or non-resonant manner, leading to different spectra of waves and in turn different energy dependence of the diffusion coefficient. Diffusion in self-generated waves is intrinsically non linear, in that the spectrum of particles is related to the diffusion coefficient, which is in turn determined by the particle spectrum and density. Non linear transport is of utmost importance for diffusive shock acceleration (see \cite{BlasiRev2013,ElenaRev} for recent reviews), in that in its absence the maximum energy of particles accelerated at SNR shocks is exceedingly low. Non linear transport may however also be crucial to shape the spectrum of CRs on Galactic scales, close to the acceleration sites of CRs or during CR escape from their host galaxies. In this article I discussed these three cases and the possible phenomenological implications that may follow. 

In the Galaxy, it is the large scale gradient of CRs that is responsible for the excitation of resonant streaming instability. The interplay between scattering on self-generated waves and 
on pre-existing, perhaps Kolmogorov like turbulence, may induce spectral breaks in the observed CR spectra \cite{BlasiAmatoSerpico,AloisioBlasi,AloisioBlasiSerpico,Carmelo}, which compare well with the features recently observed by AMS-02 and previously by PAMELA. Although this is not the only explanation of such features (see for instance \cite{Tomassetti}), it is definitely one with a well established physical foundation. A series of implications for secondary nuclei and secondary-to-primary ratios have been widely investigated in recent years. The main difficulty in testing this picture in a reliable way is that the AMS-02 data on nuclei show many anomalous behaviours that require many adjustments, not all well motivated (see for instance \cite{CarmeloNuclei}). Discriminating among models responsible for the spectral breaks is then rather difficult at present.

The rate of growth of the resonant streaming instability is faster when the CR density and its gradients are more prominent. This is the case close to sources of CRs, such as SNRs, on distance scales comparable with the coherence length of the local Galactic magnetic field in the near source region, typically $\sim 50-100$ pc. The growth of magnetic perturbations in this situation was recently investigated in Refs. \cite{Plesser,Malkov,Dangelo,Nava,NavaRecchia} for the cases of fully and partially ionized interstellar medium around the source. 
In the fully ionized case, the growth of waves resonant with particles escaped from the SNR causes an enhancement of the confinement time in the near source region, up to several hundred thousand years at GeV energies. The effect becomes negligible at energies $\gtrsim TeV$. The presence of a small amount of neutral hydrogen leads to the onset of ion-neutral damping that limits considerably the growth of waves. The fully ionized gas is expected to have density $\sim 10^{-2}-10^{-3}~\rm cm^{-3}$, hence the enhanced confinement time does not lead to a relevant grammage accumulated near the source. On the other hand, the occasional presence of a dense localized gas cloud in the near source region may change this conclusion, while not affecting the confinement time. 

The issue of the grammage accumulated in the near source regions may have important implications for the overall picture of CR transport in the Galaxy, as recently discussed by \cite{Cowsik}, although not in connection with non-linear transport near sources. The near source grammage also leads to appreciable effects on the diffuse gamma ray emission in the Galactic disc, which may be dominated, at least along some lines of sight, by the contribution of $pp$ interactions in the near source regions \cite{DangeloGamma}.  

When the electric current transported by CRs becomes large enough (Eq. \ref{eq:nr}), a non resonant branch of the streaming instability gets excited, characterized by a large growth rate \cite{Bell2004}. While this instability has been mainly invoked close SNR shocks, where CRs are expected to be accelerated, its applicability is wider. In this article I discussed the growth of the non resonant instability induced by the current of CRs escaping our Galaxy, following the investigation of Ref. \cite{BlasiEscape}. In this situation, the growth occurs on time scales of the order of years and the magnetic field grows to the level of $\sim 2\times 10^{-8}$ G. Particles that, in the absence of this effect, would be streaming freely, instead start diffusing by scattering on the self-generated waves and create a density gradient (a force) that sets the circumgalactic medium in motion, with a speed of the order of the Alfv\'en speed in the amplified field (typically $10-100$ km/s). This chain of events causes CRs to be advected away from our Galaxy and to develop an overdensity in a region of the order of a few galactic disc radii, several tens of kpc. In this region the occasional interactions of accumulated CRs with the gas in the circumgalactic medium leads to the production of neutrinos (through the decay of charged pions) and gamma rays (through the decay of neutral pions). The neutrino flux compares well with the observations of IceCube, provided there is an overdensity of $\sim 100-200$ in the gas, compatible with what is expected inside the virial radius of a structure such as our Galaxy. Interestingly, a diffuse gamma ray emission has recently been measured by Fermi-LAT from a large region around the Andromeda galaxy \cite{M31}. Whether such emission is due to the phenomenon described here or to a CR driven wind (see \S \ref{sec:GalTra}), or to some yet unknown phenomenon remains to be seen, based on additional investigation that is currently ongoing. 

\acknowledgments{The author is grateful to R. Aloisio, E. Amato, P. Cristofari, C. Evoli and O. Pezzi for useful discussions and comments. This work was partially funded through Grants ASI/INAF n. 2017-14-H.O.}




\externalbibliography{yes}
\bibliography{CR}



\end{document}